# ELM mitigation via rotating resonant magnetic perturbations on MAST

AJ Thornton[*,a], A Kirk[a], P Cahyna[b], IT Chapman[a], G Fishpool[a], JR Harrison[a], YQ Liu[a], L Kripner[b], M Peterka[b] and the MAST Team

[a]CCFE, Culham Science Centre, Abingdon, Oxon, OX14 3DB, UK

[b]Institute of Plasma Physics AS CR v.v.i, Prague, CZ

**Abstract**

The application of resonant magnetic perturbations (RMPs) produces splitting of the divertor strike point due to the interaction of the RMP field and the plasma field. The application of a rotating RMP field causes the strike point splitting to rotate, distributing the particle and heat flux evenly over the divertor. The RMP coils in MAST have been used to generate a rotating perturbation with a toroidal mode number n=3. The ELM frequency is doubled with the application of the RMP rotating field, whilst maintaining the H mode. During mitigation, the ELM peak heat flux is seen to be reduced by 50% for a halving in the ELM energy and motion of the strike point, consistent with the rotation of the applied RMP field, is seen using high spatial resolution (1.5mm at the target) heat flux profiles measured using infrared (IR) thermography.

## 1. Introduction

Edge localised modes (ELMs) are a concern for future devices due to the sudden, and repetitive heat fluxes that they deposit onto the divertor surfaces [1]. In future devices, such as ITER, the ELM heat fluxes present a limit to the operational lifetime of the divertor and therefore must be controlled. One form of ELM control is the use of resonant magnetic perturbations (RMPs) which act to perturb the edge of the plasma. ELM control can take two forms; ELM suppression occurs when the ELMs are completely eliminated by the RMPs and ELM mitigation where the ELM frequency is increased as a result of the application of the RMP. The application of non-axisymmetric RMP field causes the divertor strike point to split into lobes [2]. These lobes produce striations in the heat and particle flux at the divertor which vary toroidally and could lead to uneven erosion and deposition across the divertor [3]. The resulting uneven surface has implications for power handling in the divertor, as it exposes the divertor to large heat loads due to misalignment of the surface with the magnetic field. Rotating the applied RMP field causes the strike point splitting pattern to rotate, ensuring the divertor surface is uniformly eroded. The results presented in this paper will concentrate on the heat flux mitigation and the effect of the RMP on the strike point splitting at the divertor. The use of rotating RMP fields to change the strike point structure has been seen to be effective in the past [4, 5] which motivates this study. It should be noted that material migration will depend not only on the steady state heat loads investigated in this paper, but also on the ELM induced transient heat and particle load. Indeed, the ELM transient load may dominate the effect on material migration. To estimate the material migration, an analysis of the particle flux reduction would be required, which is beyond the scope of this paper, as a result we concentrate on the heat flux mitigation.

## 2. Rotating RMP on MAST

The Mega Amp Spherical Tokamak (MAST) is equipped with a set of internal RMP coils for ELM control [6]. Twelve lower coils, equally spaced toroidally around the vessel allow a rotating RMP with a toroidal mode number n=3 to be generated and applied to the plasma. Figure 1 a) shows the current applied to the coils to generate a rotating RMP with a single n=3 field requiring the use of 6 coils. To rotate the field, current is transferred from the initial phase of 6 coils to the final phase of 6 coils. The superposition these two coil sets then produces the rotating field. The power supplies used to generate the rotating RMP allow phase of the applied field can be rotated toroidally by ±30 degrees during the discharge. The total RMP current, $I_{RMP} = \sqrt{I_{phase1}^2 + I_{phase2}^2}$, is kept constant which gives a linear rotation in toroidal angle with time. The time evolution of the currents in the two phases of the RMP coils is shown in Figure 1 b). The rotating RMP is applied to a 400 kA lower single null (LSN), beam heated (3.5 MW), type I ELMy H mode plasma which has been seen to give ELM mitigation with the application of a RMP due to the safety factor profile being resonant with the coil field [2]. The magnitude of $I_{RMP}$ is limited in the LSN configuration for the n=3 RMP as the applied RMP generates significant core rotating braking which can lead to a transition from H to L mode and potentially the termination of the plasma [8].

The experiments reported here concentrate on the inter-ELM heat flux profiles and the peak ELM heat fluxes, previous investigations of the splitting of the ELM heat flux have been made on MAST [7]. Future work will investigate the ELM splitting with the higher resolution infrared (IR) data. The spatial resolution of the IR gives approximately 20 points in the fall off length of the RMP off heat flux profiles when extrapolated to the midplane and approximately 150 points in the profiles at the divertor covering the region where the splitting is observed.

**3. Effect of mitigation on ELMs**

Rotation of the RMP to control the strike point splitting must also continue to provide mitigation of the heat flux to the divertor seen with the static field. In a static RMP configuration, it is possible to generate n=3, 4 and 6 perturbations which produce differing levels of mitigation through an increase in the ELM frequency and a decrease in the ELM energy loss.

Figure 2 a) shows the energy loss per ELM, $\Delta W_{ELM}$, as determined by high temporal resolution EFIT reconstruction and the ELM frequency, $f_{ELM}$, for static RMP toroidal mode numbers of n=3, 4 and 6 and for rotating n=3 RMP. The product of the energy loss per ELM and the ELM frequency is seen to be constant which is consistent with previous results [8, 9]. Hence, the RMPs produce more frequent, but lower energy ELMs. The ELM frequency is increased relative to the unmitigated ELMs, with the largest increases in ELM frequency coming from the higher toroidal mode numbers. The lower levels of rotation braking seen in the n=4 and 6 RMP permit a larger RMP field to be applied leading to higher ELM frequencies. The unmitigated ELM frequency is approximately 60 Hz. Application of either a static or a rotating n=3 RMP results in the ELM frequency increasing to 120 Hz. The increase in ELM frequency with the n=3 RMP is maintained when the field is rotated in either direction.

The reduced energy loss per ELM observed in mitigated ELMs leads to a reduction in the peak heat load to the divertor, q. Infrared (IR) thermography has been used to measure the heat flux to the divertor at a temporal resolution of 200 $\mu s$ as a function of $\Delta W_{ELM}$, as shown in Figure 2 b) or static n=4 and 6 RMP and both static and rotating n=3. Mitigation is seen to be effective at reducing the ELM heat flux, with a 2 fold reduction in $\Delta W_{ELM}$ resulting in a 44% reduction of the divertor peak heat flux. The reduction seen in the discharges reported here is larger than that seen previously in 600 kA LSN plasmas [10]. The n=3 mitigated heat flux is of the order 13 MW/m$^2$ (compared to unmitigated levels of up to 23 MW/m$^2$), and the level of mitigation seen when the field is static or rotating is the same, indicating that a rotating RMP is as effective at mitigating the heat flux as a static field, as required for ITER.

The peak heat flux to the divertor is determined by the energy deposited by the ELM, the duration of the ELM and the wetted area over which the energy is deposited. These additional parameters are not considered when the reduction in peak heat flux is considered alone. In order to account for the changes in area during the ELM, the heat flux factor can be used which is the energy of the ELM normalised by the wetted area and the square root of the ELM duration [11]. Naturally, the inclusion of additional parameters in the calculation compared to the peak heat flux increase the uncertainty in the calculated values. However, the heat flux factor (calculated using the method in [11]) is seen to reduce in the ELMs studied here, falling from 0.31 MJ m$^{-2}$ s$^{-0.5}$ to 0.15 MJ m$^{-2}$ s$^{-0.5}$ for a halving of the ELM energy, consistent with the results seen in the peak heat flux analysis.

## 4. Measurements of the strike point splitting

Heat flux profiles across the divertor can be used to determine the level of strike point splitting generated by the RMP. The strike point in MAST naturally sweeps outwards in radius during the discharge due to the fringing field of the central solenoid. Figure 3 shows the heat flux profile to the divertor as a function of time and distance from the last closed flux surface (LCFS), $\Delta R_{LCFS}$, for an anticlockwise rotation of the RMP field. The location of the strike point is converted to $\Delta R_{LCFS}$ to remove the strike point sweeping. The diagonal bands running from top left to bottom right are a result of localised hotspots at fixed radius on the divertor surface. The IR profiles are smoothed and averaged in space to minimise the effect of hotspots. The splitting of the strike point into three lobes can be seen in Figure 3, with the distance between the lobes being largest at the start of the rotating period and decreasing as the field is rotated. Individual measured IR profiles extracted from Figure 3 are shown in Figure 4 for profiles at the start, middle and end of the rotation. The IR data supports the modelling which shows the lobe separation decreasing in time. There is some evidence of broadening of the lobe at $\Delta R_{LCFS} = 0.03$m in the middle rotation phase (red curve), though the resolution and signal to noise of the data prevent this observation from being conclusive.

## 5. Modelling of the splitting

In order to interpret the IR data in Figure 3 and Figure 4, modelling of the heat flux deposited onto the divertor can be performed using field line tracing. Field lines are traced through the vacuum magnetic field (RMP field, error field and error field correction) in order to map given locations on the divertor to the deepest point reached inside the plasma in terms of normalised flux, $\Psi_{min}$. Using the technique developed by Cahyna et al [Cahyna2013], the modelled strike point heat flux is calculated using the $\Psi_{min}$ profile at the location of the IR camera, converted to a heat flux by assuming an exponential decay over a decay length, $\lambda_q$. The heat flux profile at the divertor is thought to be generated by an exponential fall off across the scrape off layer (SOL) and Gaussian diffusion along the divertor leg [13]. Consequently, the exponential fall off is then convolved with a Gaussian of width, S. The $\lambda_q$ and S values used are determined via fitting to the measured heat flux profiles without the application of the RMP. In Figure 4 the measured IR profile at the end of the rotation (blue line) is compared to the modelled profile (green line) and shows good agreement in lobe location. The modelled profile is reduced in amplitude by 50% compared to the measured for clarity, but differences in the lobe magnitude for the outer lobes can be seen. It should be noted that the model places the LCFS heat flux at any point where $\Psi_{min} < 1$ and there is no effect of screening or plasma response included. Whilst the locations of the lobes agree with the modelling, the differences in the magnitude of the modelled profile and the measured IR profile suggest that there is screening from the plasma. The inclusion of the plasma response has been seen to affect penetration of the field lines into the plasma, thereby decreasing the magnitude of the modelled profile in line with the measured profiles [14]. Previous modelling on MAST [7] has also shown that there is evidence for screening of the applied RMP in MAST discharges which is consistent with the result presented in this paper.

In the case of a static, non-rotating RMP field, the strike point splitting pattern varies toroidally, with the pattern seen on the IR camera determined by the toroidal location of the camera. If the rotating field was to be generated by physically rotating the RMP coils, then the rotation would be equivalent to moving the IR camera in the static RMP field and a smooth motion of the strike point position would result. However, there are two effects which come into play in the case of a rotating RMP field generated by superposing the two RMP coil sets. Firstly, the superposition would be equivalent to a purely rotated n=3 field, but due to the proximity of coils there are higher harmonics present. These near field effects of the coil produce a broadening of the heat flux as the RMP is rotated, as opposed to a smooth motion of the lobes inwards, as can be seen in Figure 5 a) for anticlockwise rotation of the field. The second effect arises from the evolution of the safety factor in the plasma, which causes the separation of the strike point splitting to decrease with increasing time through the discharge with a static RMP field. The modelled heat flux including both of these effects is shown in Figure 5 b) and shows good qualitative agreement with the measured IR structure in Figure 3. In MAST, the modelling suggests that the motion of the strike point splitting during rotation of the RMP has a component from the safety factor and from the rotation. A comparison of the effect on the strike point pattern between the rotation of the RMP and the safety factor evolution can be seen using IR measurements of a clockwise rotation of the RMP, shown in Figure 6. In the clockwise rotation (as opposed to the anticlockwise rotation shown in Figure 4), the motion of the splitting from RMP field and the safety factor will oppose one another. Comparison of the three profiles in Figure 6 shows that there is little motion of the strike point splitting, unlike in Figure 4 when the RMP motion reinforces the safety factor. It is therefore the case that the safety factor evolution and the rotation of the RMP are similar in magnitude, but that rotation of the RMP does affect the strike point splitting pattern as required.

## 6. Conclusion

Experiments have shown that ELM control using a rotating RMP with a toroidal mode number of n=3 can generate ELM mitigation and is successful at producing motion of the strike point splitting. Application of the rotating RMP doubles the ELM frequency relative to the unmitigated case, producing mitigated ELMs with a frequency of 120 Hz. The increase in the ELM frequency is limited by the size of the RMP field which can be applied as a result of core rotation braking in the case of an n=3 RMP. Larger increases in ELM frequency have been generated with higher toroidal mode number RMP where the braking is lower. Successful mitigation of the ELM heat flux has been produced with the rotating RMP, with the peak falling from 23 MW/m$^2$ to 13 MW/m$^2$ (carbon plasma facing materials are used in MAST) for a halving of the energy loss per ELM which consistent with the static RMP reduction. A similar halving of the heat flux factor is also seen between mitigated and unmitigated ELMs on MAST. Measurements of the strike point splitting have shown that the application of an n=3 rotating field leads to splitting of the strike point and this strike point splitting varies with the phase of the rotating field. Qualitative agreement has been seen between the evolution of the measured strike point splitting and modelling which both show a consistent decrease in the separation of the lobes when the RMP is rotated anticlockwise, coupled with the motion from the safety factor changing. These data support the fact that a rotating RMP is an effective means of manipulating the strike point splitting in RMP applied discharges. The effectiveness of the rotating RMP is likely to be enhanced in steady state devices where the q remains constant and when a full rotation of the RMP field can be generated. The results on MAST are promising for ITER. It should be noted that extrapolation from MAST to ITER has several caveats. Firstly, the close fitting vessel wall on ITER means that interaction of the outer lobes with the main chamber are more likely than on MAST. In MAST, there is no interaction of the lobes with the main chamber wall, only with the divertor PFCs due to the remote vessel wall that is part of the MAST design. Secondly, ITER will operate in a semi-detached divertor regime, such a regime has not been investigated on MAST.

**Acknowledgements**

This work was part-funded by the RCUK Energy Programme [grant number EP/I501045] and by the European Union's Horizon 2020 research and innovation programme. To obtain further information on the data and models underlying this paper please contact PublicationsManger@ccfe.ac.uk. The view and opinions expressed herein do not necessarily reflect those of the European Commission. This work was also part-funded by the Grant Agency of the Czech Republic under grant P205/11/2341.

**Figure captions**

Figure 1 a). Applied RMP field for the starting phase and the ending phase for a 30 degree clockwise rotation of the applied field. Positive coil current represents a radial B field outwards from the plasma. b) Current applied to each phase of the RMP during the rotation of the field.

Figure 2. a) The energy lost per ELM, $\Delta W_{ELM}$, versus the ELM frequency, $f_{ELM}$, for a RMP with a toroidal mode numbers, n=3 static and rotating (red diamond), static n=4 (blue triangles) and static n=6 (green squares) and for RMP off cases (black circles). b) The peak heat flux at the divertor, q, as a function of the energy loss per ELM, $\Delta W_{ELM}$, for the same range of toroidal mode numbers as presented in a).

Figure 3. The heat flux at the outer divertor as measured by IR thermography. The time is shown along the x axis and the distance from the last closed flux surface (LCFS) is shown on the y axis. The grey vertical bars are the locations of the ELMs in the discharge.

Figure 4. Measured IR profile at the start of the rotation (black line), half way through the rotation (red line) and at the end of the rotation (blue line), for anticlockwise rotation. The green line shows modelling of the profile (reduced by 50% in scale) to match with the end of the rotation (blue line). The RMP off profile is shown as the dashed line.

Figure 5. Modelled heat flux pattern using vacuum fields, for a) constant safety factor of the plasma and rotating RMP field and b) varying safety factor and rotating RMP. RMP rotation starts at 0.48s in both cases, and in the anticlockwise direction.

Figure 6. Measured IR profile at the start of the rotation (black line), half way through the rotation (red line) and at the end of the rotation (blue line), for clockwise rotation.

# Figures

## Figure 1

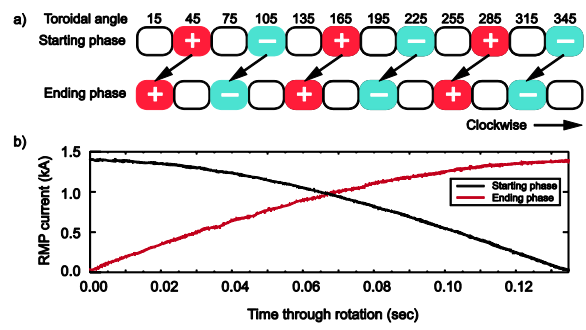

**Figure 2**

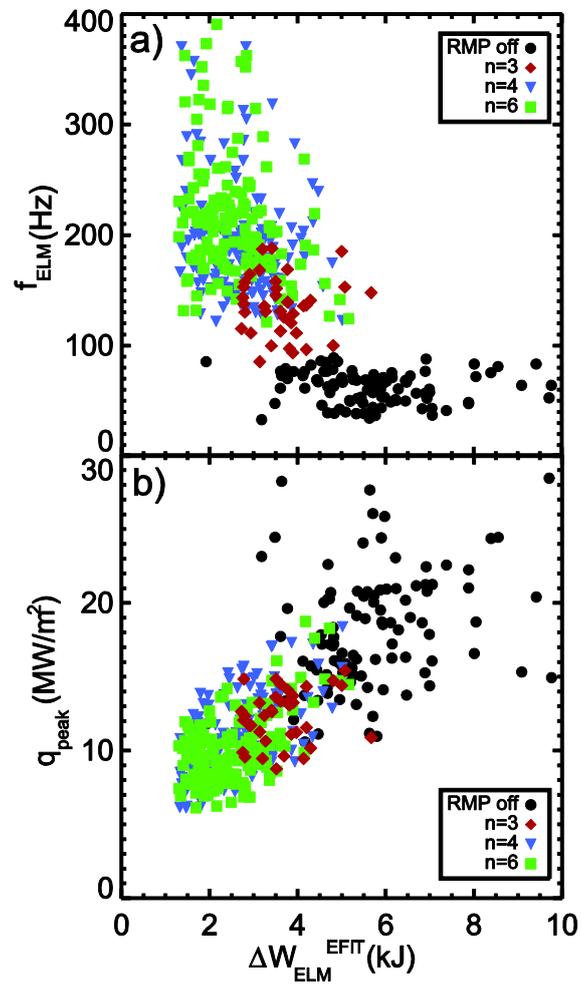

**Figure 3**

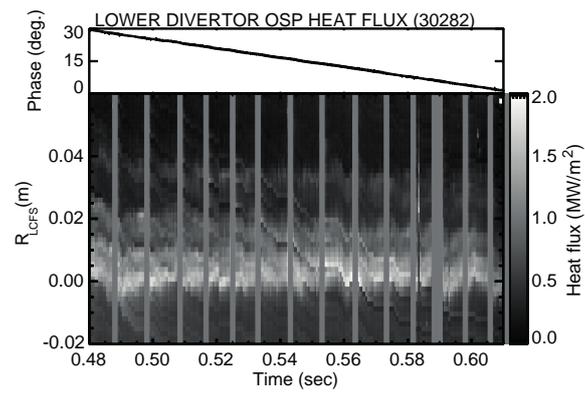

**Figure 4**

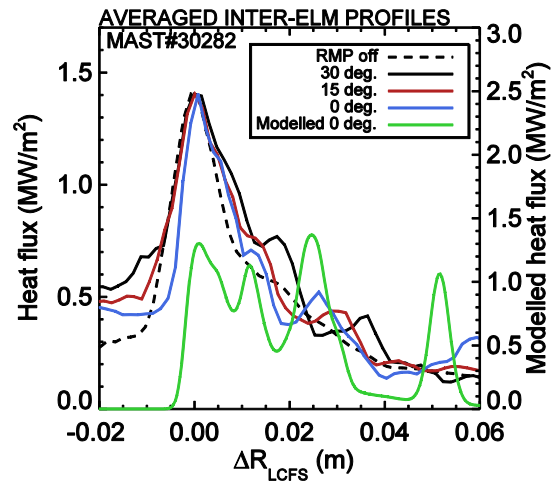

**Figure 5**

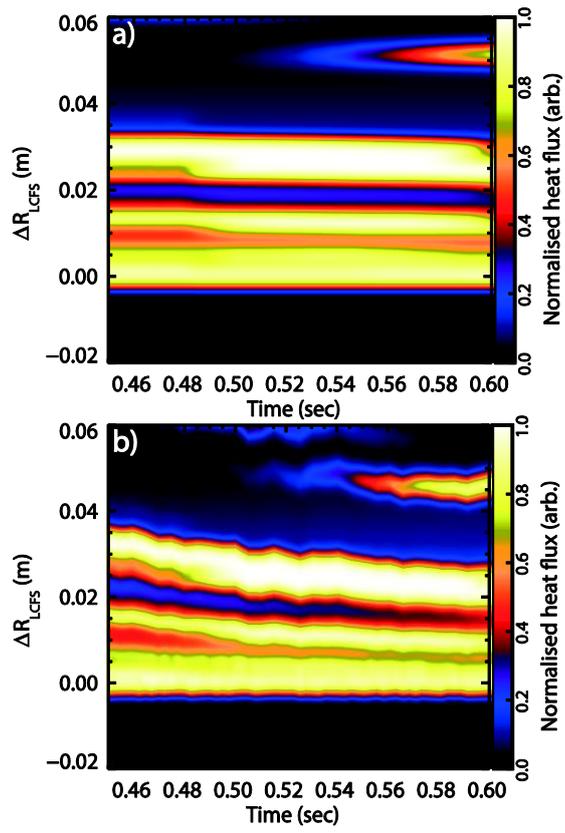

**Figure 6**

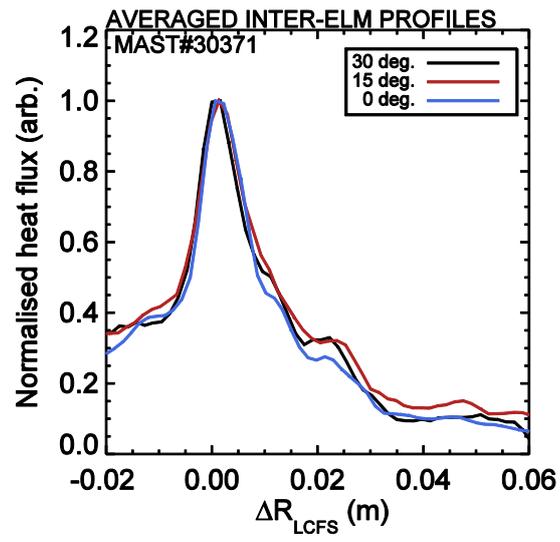